\documentclass[conference]{IEEEtran}
\IEEEoverridecommandlockouts

\usepackage{cite}
\usepackage{amsmath,amssymb,amsfonts}
\usepackage{graphicx}
\usepackage{textcomp}
\usepackage{xcolor}
\usepackage{multirow}
\usepackage{subcaption}
\usepackage{braket}
\usepackage{algorithm}
\usepackage{algpseudocode}  
\usepackage{amsmath}        
\def\BibTeX{{\rm B\kern-.05em{\sc i\kern-.025em b}\kern-.08em
    T\kern-.1667em\lower.7ex\hbox{E}\kern-.125emX}}
\begin{document}

\title{HOPPS: Hardware-Aware Optimal Phase Polynomial Synthesis with Blockwise Optimization for Quantum Circuits\\
}

\author{\IEEEauthorblockN{Xinpeng Li}
\IEEEauthorblockA{\textit{Dept. of Computer and Data Science} \\
\textit{Case Western Reserve University}\\
Cleveland, US \\
xxl1337@case.edu}
\and
\IEEEauthorblockN{Ji Liu}
\IEEEauthorblockA{\textit{Mathematics and Computer Science Division} \\
\textit{Argonne National Laboratory}\\
Lemont, US \\
ji.liu@anl.gov}
\and
\IEEEauthorblockN{Shuai Xu}
\IEEEauthorblockA{\textit{Dept. of Computer and Data Science} \\
\textit{Case Western Reserve University}\\
Cleveland, US  \\
ssx214@case.edu}
\and
\IEEEauthorblockN{Paul Hovland}
\IEEEauthorblockA{\textit{Mathematics and Computer Science Division} \\
\textit{Argonne National Laboratory}\\
Lemont, US \\
hovland@anl.gov}
\and
\IEEEauthorblockN{Vipin Chaudhary}
\IEEEauthorblockA{\textit{Dept. of Computer and Data Science} \\
\textit{Case Western Reserve University}\\
Cleveland, US \\
vxc204@case.edu}

}

\maketitle

\begin{abstract}

Blocks composed of \(\{ \text{CNOT}, R_z \}\) are ubiquitous in modern quantum applications, notably in circuits such as QAOA ansatzes and quantum adders. After compilation, many of them exhibit large CNOT counts or depths, which lowers fidelity. Therefore, we introduce HOPPS: a SAT-based hardware-aware optimal phase polynomial synthesis algorithm that could generate \(\{ \text{CNOT}, R_z \}\) blocks with CNOT count or depth optimality.

Sometime \(\{ \text{CNOT}, R_z \}\) blocks are large, such as in QAOA ansatzes, HOPPS’s pursuit of optimality limits its scalability. To address this issue, we introduce an iterative blockwise optimization strategy: large circuits are partitioned into smaller blocks, each block is optimally refined, and the process is repeated for several iterations.

Empirical results show that HOPPS is more efficient comparing with existing near-optimal synthesis tools. Used as a peephole optimizer, HOPPS reduces the CNOT count by up to \(50.0\%\) and the CNOT depth by up to \(57.1\%\) under OLSQ. For large QAOA circuit, after mapping by Qiskit, circuit can be reduced CNOT count and depth by up to \(44.4\%\) and \(42.4\%\) by our iterative blockwise optimization.
\end{abstract}


\section{Introduction}
NISQ quantum devices, both near-term devices~\cite{ai2024quantum-willow, mckay2023benchmarking-heron} and long-term scalable quantum architectures~\cite{monroe2014large-QCCD, bluvstein2022quantum-FPQA}, are constrained by limited qubit connectivity, permitting two-qubit gates only between physically connected qubits. Additionally, two-qubit gates tend to have higher error rates and longer durations. As a result, increasing their count or depth reduces overall circuit fidelity. Therefore, optimizing the number and depth of two-qubit operations is necessary for NISQ quantum devices \cite{yan2024quantum, kusyk2021survey}.

Among quantum applications, circuits composed of \(\{ \text{CNOT}, R_z(\theta) \}\) gates are frequently used in designs such as the Quantum Approximate Optimization Algorithm (QAOA)~\cite{farhi2014quantum}, quantum adders~\cite{orts2020review}, and the quantum Fourier transform (QFT)~\cite{weinstein2001implementation}. Fortunately, such circuits can be efficiently encoded as low-degree polynomials over the finite field \( \mathbb{F}_2 \) using a phase polynomial representation~\cite{montanaro2016quantum, dawson2004quantum}. Unlike unitary matrix representations, which require \( O(2^{2n}) \) memory to store, the phase polynomial representation scales linearly with the number of qubits, i.e., \( O(n^2) \). This compact representation facilitates both phase polynomial circuit synthesis \cite{vandaele2022phase, de2020architecture, meuli2018sat} and circuit verification \cite{peham2023equivalence}.

In this work, we focus on phase polynomial circuit synthesis, which is the process of generating a \(\{ \text{CNOT}, R_z(\theta) \}\) circuit from its phase polynomial representation. We observe that existing studies have explored both heuristic and optimal solutions for logical-level circuits \cite{vandaele2022phase, meuli2018sat}, as well as heuristic methods for physical-level circuits \cite{de2020architecture, montanez2025optimizing, dreier2025connectivity}. However optimal synthesis under arbitrary hardware constraints remains unaddressed. 
This motivates us to introduce an algorithm for Hardware-Aware Optimal Phase Polynomial Synthesis, called \textbf{HOPPS}.

Because phase polynomial synthesis operates entirely in a binary encoding, SAT is well-suited to circuit synthesis. On NISQ hardware, CNOT gates are error-prone and have relatively long durations comparing with \(R_z(\theta)\), we provide both CNOT count and depth optimal synthesis results. We claims HOPPS can also provide doubly optimal based on one of metrics. For example, a CNOT-based doubly optimal solution not only minimizes the CNOT count but also achieves the minimal CNOT depth under that optimal CNOT count.

While HOPPS provides optimal physical circuits, it faces scalability challenges. Therefore, we explore how optimal solutions can be practically leveraged. One key application of HOPPS is peephole optimization on mapped circuits. However, in some cases—such as QAOA circuits, which are almost entirely composed of \(\{ \text{CNOT}, R_z(\theta) \}\) gates—the block may be too large for direct optimization. To address this, we propose an \textbf{Iterative Blockwise Optimization} approach: repeating (1) partitioning circuit into smaller blocks (2) optimizing blocks using HOPPS. Fortunately, the independence of blocks enables efficient \textbf{parallel execution} for each iteration.

Our main contributions are summarized as follows:

\begin{itemize}

    \item We propose a SAT-based algorithm, named \text{HOPPS}, that synthesizes doubly optimal CNOT count or CNOT depth \(\{ \text{CNOT}, R_z(\theta) \}\) circuits under hardware topology.

    \item We introduce an iterative blockwise optimization strategy for large-scale \(\{ \text{CNOT}, R_z(\theta) \}\) circuits.
    
    \item We propose a parallel framework that enables iterative blockwise optimization to run efficiently on HPC systems.

\end{itemize}

The remainder of this paper is organized as follows. Section~\ref{sec:background} provides the necessary background. In Section~\ref{sec: Problem} we explain the synthesis problem and our objective. Section~\ref{sec:sat-model} presents our SAT-based synthesis algorithm HOPPS. Section~\ref{sec: Circuit Optimization} presents the peephole optimization, iterative blockwise optimization strategies and their parallel implementation. Section~\ref{sec: Related Work} presents some related work. Experimental results are reported in Section~\ref{sec:experiments}. Finally, Section \ref{sec: Discussion} discusses the benefits of using HOPPS, and Section \ref{sec: Conclusion} summarizes our findings and outlines directions for future work.

\begin{figure*}[ht]
    \centering
    \includegraphics[width=\textwidth]{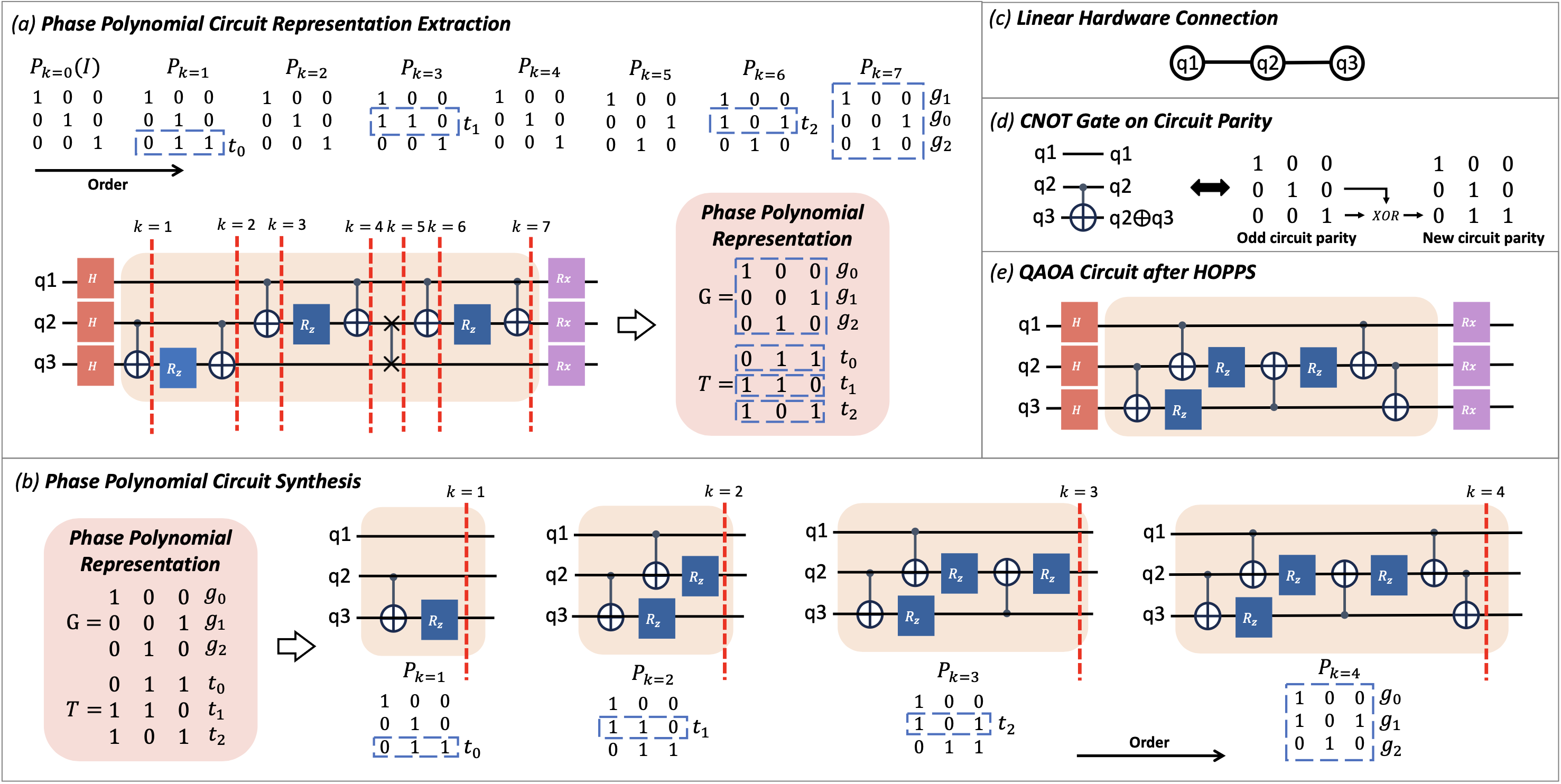}  
\caption{Explanation of Phase Polynomial Representation and Synthesis. 
\textbf{(a)} Extracting the phase polynomial representation from a QAOA circuit. The circuit is scanned from beginning to end, starting with an identity matrix. Each CNOT gate updates the circuit’s parity matrix. When an \(R_z(\theta)\) gate is encountered, the parity vector associated with that qubit is recorded and stored in the table \(T\). After processing the entire circuit, only the final circuit parity matrix \(G\) and the table \(T\) are retained.
\textbf{(b)} Synthesizing a circuit from the phase polynomial representation. CNOT gates are iteratively added, and the parity matrix is updated accordingly. When a qubit’s parity vector matches one in \(T\), an \(R_z(\theta)\) gate is inserted. After all \(t \in T\) are found, we continuously add CNOT gates to make the parity match \(G\). \textit{Note: This process is not used in this paper; it is provided solely to aid in understanding the problem.} \textbf{(c)} Linear hardware topology. The circuits in (a), (b), and (e) obey this constraint, meaning that qubit 1 and qubit 3 cannot be directly connected.
\textbf{(d)} Effect of a CNOT gate on circuit parity. A CNOT operation updates the parity matrix by applying a bitwise \texttt{XOR} between the control and target vectors, replacing the control vector with the result.
\textbf{(e)} Optimal solution for the QAOA circuit in (a) under the hardware topology in (c).}

    \label{fig: Combined}
\end{figure*}

\section{Background}
\label{sec:background}

\subsection{Phase Polynomial Representation}
The Phase Polynomial Representation is an alternative representation for a quantum circuit composed exclusively of gates from the set \(\{ \text{CNOT}, R_z(\theta) \}\). When an \(n\)-qubit circuit is restricted to these gates, it can be expressed by using the circuit-polynomial correspondence \cite{montanaro2016quantum}, consisting of a phase polynomial function \(p(\cdot)\) and a linear reversible function \(g(\cdot)\). The action on a basis state \(\ket{x}\) can be written as
\begin{equation}
\ket{x} \mapsto e^{2\pi i p(x)} \ket{g(x)}.
\label{eq:phase-polynomial-mapping}
\end{equation}

Here, the phase polynomial is given by
\begin{equation}
p(x) = \sum_{i=0}^{2^n-1} \theta_i f_i(x).
\label{eq:phase-polynomial}
\end{equation}
First, in \(p(x)\), each \(f_i(x)\) is a linear Boolean function,
\(
f_i: \mathbb{F}_2^n \rightarrow \mathbb{F}_2
\)
, thus can be represented as a binary coefficient vector \(\mathbf{t}^{(i)} = (t^{(i)}_0, t^{(i)}_1, \dots, t^{(i)}_n) \in \mathbb{F}_2^n\) and written as 
\begin{equation}
f_i(x) = \mathbf{t}^{(i)} \cdot x = t^{(i)}_0 x_0 \oplus t^{(i)}_1 x_1 \oplus \cdots \oplus t^{(i)}_n x_n .
\label{eq:fi-boolean-function}
\end{equation}

We collect all these vectors from \(f_0,\dots,f_{n-1}\) form a set of binary vectors \(T = \{\mathbf{t}^{(0)}, \dots, \mathbf{t}^{(n-1)}\}\), which is also referred to as the ``parity table'' \cite{vandaele2022phase}. In the Eq.~\ref{eq:phase-polynomial}, each boolean function \(f_i(x)\) is associated with a corresponding parameter \(\theta_i\), collectively forming the parameter set \( \Theta = \{\theta_0, \dots, \theta_{n-1} \} \). The sets \(\Theta \) and \(T\) are in one-to-one correspondence.

Second, the function \(g(x)\), \(
g: \mathbb{F}_2^n \rightarrow \mathbb{F}_2^n
\), in the Eq.~\ref{eq:phase-polynomial-mapping} can also be expressed as a vector-valued function \(g(x) = (g_0(x), g_1(x), \dots, g_{n-1}(x))\), where each component \(g_j(x)\) is a linear Boolean function. Each \(g_j(x)\) is defined by a coefficient vector \(\mathbf{g}^{(j)} = (g^{(j)}_0, g^{(j)}_1, \dots, g^{(j)}_n)\) and is written as
\begin{equation}
g_j(x) = \mathbf{g}^{(j)} \cdot x = g^{(j)}_0 x_0 \oplus g^{(j)}_1 x_1 \oplus \cdots \oplus g^{(j)}_n x_n .
\label{eq:gj-boolean-function}
\end{equation}

Similarly, the linear reversible function 
\(g(x)\) can be represented by a binary matrix \(G\), where each row corresponds to the coefficient vector \(\mathbf{g}^{(j)}\) of the Boolean function \(g_j(x)\). That is, the transformation \(g(x)\) is realized through matrix-vector multiplication:
\[
g(x) = Gx \mod 2,
\]
where \(G \in \mathbb{F}_2^{n \times n}\) is an invertible binary matrix, also known as a parity matrix.

In summary, a circuit composed of gates from the set \(\{ \text{CNOT}, R_z(\theta) \}\) can be succinctly represented by a parity table \(T\), parameter set \(\Theta \) and a parity matrix \(G\).

\subsection{Phase Polynomial Circuit Representation Extraction}
\label{subsec: Phase Polynomial Circuit Representation Extraction}
Given a circuit, we do not have to extract Phase Polynomial Representation writing in Eq.~\ref{eq:phase-polynomial-mapping}. For an \(n\)-qubit circuit, it begins with an initial state  
\[
\ket{x} = \ket{x_0}\ket{x_1} \dots \ket{x_{n-1}},
\]  
where each qubit is initialized independently. 
When a CNOT is applied, it will perform:
\[
\text{CNOT}: \ket{x_i}\ket{x_j} \mapsto \ket{x_i}\ket{x_i \oplus x_j}.
\]
This process can be efficiently tracked parity of the circuit \(P\) called \textit{circuit parity}. Because \(P\) changes over time, we write \(P_{k,i,j}\) for the entry in row \(i\) and column \(j\) at time step \(k\). The initial circuit parity \(P_0\) is typically the identity matrix. A CNOT gate performs an \texttt{XOR} between two row vectors of \(P\), replacing one of the rows, as illustrated in Fig.~\ref{fig: Combined}(d). We traverse the circuit, updating the circuit parity under each CNOT; whenever we encounter an \(R_z(\theta)\) on the \(i\)th qubit, we record the current circuit parity (the \(i\)th row of the parity table) into \(T\). When the process completes, the final circuit parity is \(G\). We provide an example in 
Fig.~\ref{fig: Combined} (a).

\subsection{Phase Polynomial Synthesis}
\label{subsec: Phase Polynomial Synthesis}

Phase polynomial synthesis is the reverse process of extracting a phase polynomial representation from a circuit. It involves reconstructing a quantum circuit from: 1) a phase polynomial parity table \(T\); 2) a parameter set \(\Theta\); and 3) a linear transformation parity matrix \(G\). During synthesis, we continuously add CNOT gates to match each \(t_i(x)\) and matrix \(G\) finally. The \(R_z(\theta_i)\) gates are placed at the circuit positions corresponding to \(t_i(x)\).

\textbf{Example.}
To illustrate the synthesis process, we provide an example of a three-qubit QAOA cost layer synthesis. The parity table \(T\) and the final (target) circuit parity matrix \(G\) are
\[
T = \begin{bmatrix}
1 & 0 & 1 \\
1 & 1 & 0 \\
0 & 1 & 1
\end{bmatrix}
\qquad
G = \begin{bmatrix}
1 & 0 & 0 \\
0 & 0 & 1 \\
0 & 1 & 0
\end{bmatrix}.
\]

For convenience, we refer to each row in the parity table \(T\) as a \emph{term parity}, since it corresponds to a term in the phase polynomial \(p(x)\). 

As shown in Fig.~\ref{fig: Combined} (b), the circuit is initialized with the identity parity matrix. In order to implement the first term parity \(t_0 = [0, 1, 1]\) on qubit \(q_3\), a CNOT gate is applied with control on \(q_2\) and target on \(q_3\). This corresponds to updating \(P_{0,3,j}\) by performing a \texttt{XOR} with \(P_{0,2,j}\) if  \(P_{0,2,j} = 1\) for all \(0 \leq j < 3\). As a result, the third row of the matrix becomes equal to \(t_0\), and we can apply \(R_z\) on it.
We continue adding CNOT gates until all \(t \in T\) are applied as \(R_z(\theta)\) gates in the circuit. Note that each term parity should be synthesized only once, and the order of synthesis does not affect correctness. After finishing syshtesis \(T\), we need some CNOT gates to convert circuit parity to matrix \(G\). Finally, we assign \(\theta_i\) to their corresponding term parity positions. At this point, the synthesis process terminates. Synthesizing such circuits is an NP-complete problem\cite{amy2017cnot}. 

We emphasize that a single synthesized circuit is reused for multiple parameter sets \(\Theta\). The core of the synthesis process is to make the circuit parity match every term parity in \(T\) at least once and the final parity matrix \(G\) in the end.

\subsection{SAT}

SAT (Boolean satisfiability) employs a solver to decide if a satisfying Boolean assignment exists for a specified set of constraints. The given problem is encoded as boolean variables and boolean formulas using operators \texttt{AND}, \texttt{OR}, and \texttt{NOT} to constrain variables. In this paper, we encode phase polynomial synthesis—which can naturally be expressed in binary—as a problem over Boolean variables and constraints.

To obtain an optimal solution, the SAT solver performs incremental satisfiability checks: it starts from a known lower bound and progressively enlarges the search space. The first instance that proves satisfiable yields an optimal solution.

\subsection{Hardware Restriction and SWAP Based Mapper}
\label{subsec: Hardware Restriction}

On real Noisy Intermediate-Scale Quantum (NISQ) devices, two-qubit gates are often limited to specific pairs of physical qubits, depending on the device’s connectivity topology. We can model this topology as a graph \(CP = (Q, E)\), where each node \(q_i \in Q\) represents a physical qubit and each edge \(e \in E\) indicates that a two-qubit gate can be directly applied between the connected qubits. This graph is called the coupling map.

\begin{figure}[htbp]
    \centering
    \includegraphics[width=0.92\linewidth]{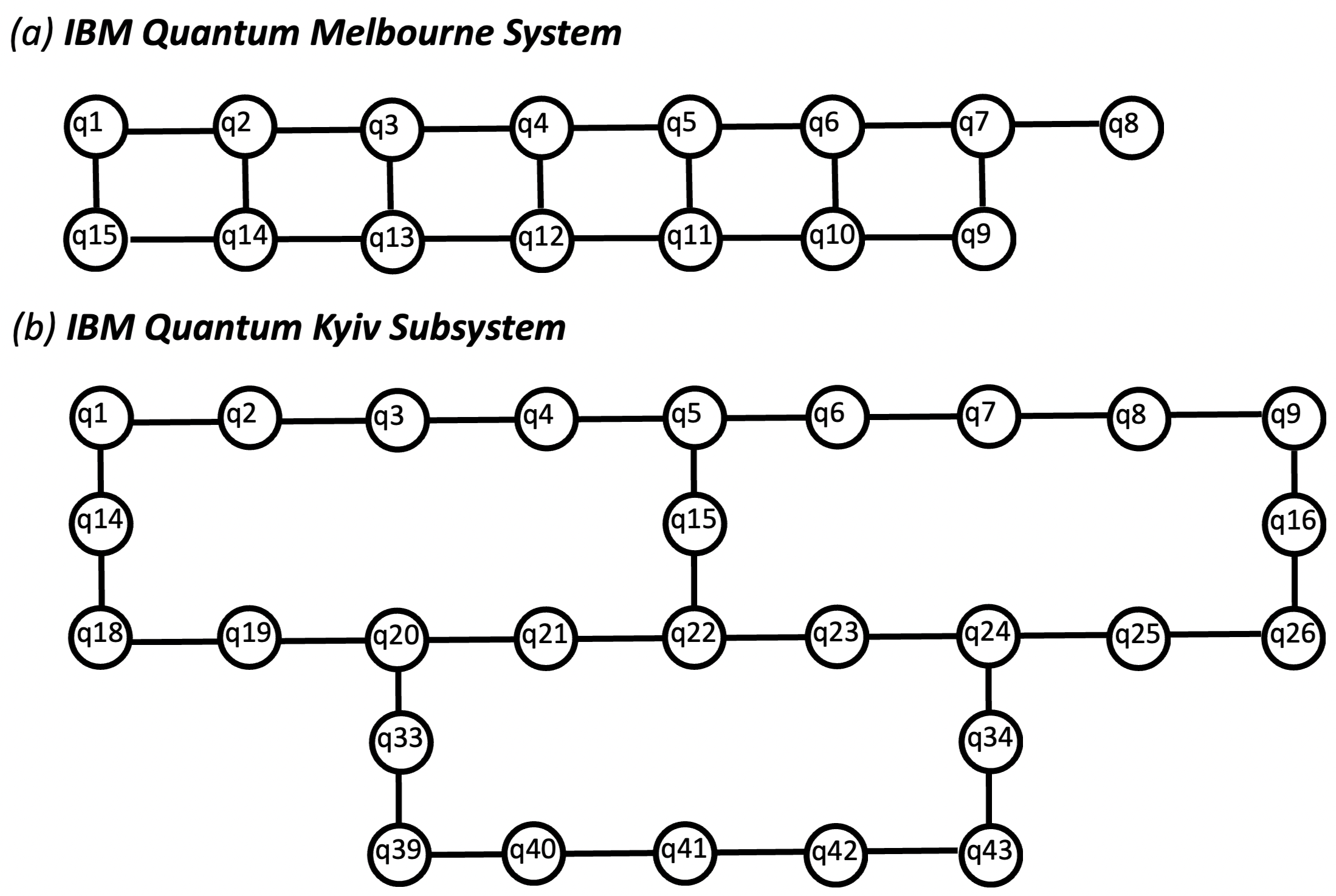}

    \caption{Overview of the coupling map (qubit-connectivity topology) for the \texttt{ibmq\_melbourne} system and \texttt{ibmq\_kyiv} subsystem. Nodes represent physical qubits, and edges indicate allowed
    two-qubit gate operations on two physical qubits.
}
    \label{fig:brisbane-topology}
\end{figure}

\textbf{Example.} Fig.~\ref{fig:brisbane-topology} shows the coupling map \(CP\) for \texttt{ibmq\_melbourne} system and \texttt{ibmq\_kyiv} subsystem. As seen in the figure, the connectivity is limited—many qubits are connected in a nearly linear or sparse topology. 

To apply a two-qubit gate between no adjacent qubit pairs, traditional compilation techniques insert SWAP gates to move qubits closer until the desired pair becomes adjacent. This process is commonly referred to as \emph{layout synthesis}~\cite{tan2020optimal} or \emph{qubit mapping}~\cite{li2019tackling}. 
SWAP operations are expensive, typically decomposing into three CNOT gates; therefore, the mapping procedure should minimize the number of required SWAPs. 

\section{Hardware-Aware Optimal Phase Polynomial Synthesis}
\label{sec: Problem}
The key novelty of our phase-polynomial synthesis is twofold: \emph{hardware-aware, optimality}. We elaborate on these objectives below.

\subsection{Hardware-Aware Synthesis}
Hardware-aware synthesis requires that inserted CNOT gates respect the hardware topology, which increases the complexity of the synthesis process. In Fig.~\ref{fig: Combined}(e), we show the result of applying hardware-aware synthesis to the QAOA circuit from Fig.~\ref{fig: Combined}(a) under a linear topology in Fig.~\ref{fig: Combined}(c).

\subsection{Optimal Synthesis}
\label{subsec: Optimality}
In quantum circuit synthesis, CNOT count and CNOT depth are two commonly used metrics, due to two-qubit gates are error-prone and have relatively long durations. However, a circuit that is optimal in terms of CNOT count is not necessarily optimal in depth, and vice versa.

Nevertheless, we propose the notion of ``doubly optimal synthesis''. This refers to cases where, under an optimal CNOT count, the CNOT depth is also minimized, and conversely, under an optimal CNOT depth, the CNOT count is also minimized. We refer to these two scenarios as \textbf{CNOT-based doubly optimal} and \textbf{depth-based doubly optimal}, respectively.

\section{HOPPS: An SAT-based \{CNOT, Rz\} Circuit Synthesis Algorithm}
\label{sec:sat-model}

The HOPPS algorithm is a SAT-based approach that takes the initial parity matrix \(I\), final parity matrix \(G\), parity table \(T\), term rotation angles \(\Theta\), and the coupling map graph \(CP\) as input. It outputs a quantum circuit \(qc\). The algorithm supports two optimization objectives: minimizing the total CNOT count (HOPPS-CNOT) or minimizing the CNOT depth (HOPPS-Depth). Both of them can be doubly optimized. Since both objectives share the same variable encoding, but differ in constraints, we first introduce the common variables and constraints, followed by a description of the differences.

\subsection{Common Variable Encoding}

The primary variables are derived from two aspects:

\begin{itemize}
    \item \textbf{Encoding CNOT gates:}  
    For each edge \(e(q_i, q_j)\) in the coupling map, two possible CNOT gates can be applied—one with \(q_i\) as the control and \(q_j\) as the target, and the other with \(q_j\) as the control and \(q_i\) as the target. We denote \(\vec{e}\) as a directed edge in the directed coupling map \(\vec{CP}\), and define a binary variable \(cnot_{k, \vec{e}}\) to indicate whether a CNOT gate is applied at time step \(k\) along the directed edge \(\vec{e}\).

    \item \textbf{Encoding circuit parity:}  
    To track how the circuit parity evolves with CNOT gates applied just like in Fig.~\ref{fig: Combined} (b), we define a three-dimensional binary matrix \(P_{k, i, j}\), where \(k\) represents the time step and \(i, j\) are row and column indices indicating the circuit parity state at that moment. At each time step \(k\), the circuit parity is updated based on the CNOT operations applied, represented by the vector \([cnot_{k, \vec{e}} \mid \vec{e} \in \vec{E}]\).
\end{itemize}

\subsection{Common Constraints}
\subsubsection{Initial Final Circuit Parity Matrix}

First, we constrain the initial and final circuit parity. Assuming the initial parity matrix is represented by matrix \(I\) and the final by matrix \(G\), the constraints are as follows.

\begin{equation}
\bigvee_{i,j} I_{i,j} = P_{0,i,j}
\label{eq:initial_parity}
\end{equation}

\begin{equation}
\bigvee_{i,j} G_{i,j} = P_{k+1,i,j}
\label{eq:final_parity}
\end{equation}

\subsubsection{Parity Table}

Second, our goal is to find all \(n_{\text{term}}\) term parities in the circuit parity. For each term parity \(t_{nt}\), there must exist at least one row in the circuit parity that matches them:

\begin{equation}
\bigvee_{nt} \exists_{k, i} \texttt{AND}( t_{nt, j} = P_{k,i,j} ).
\label{eq:parity_match}
\end{equation}
\subsubsection{CNOT Dependence}
Third, the circuit parity \(P_{k+1}\) is updated when a CNOT gate is applied at the time step \(k\). Suppose a CNOT gate is inserted with a control qubit \(q_i\) and a target qubit \(q_j\) at time \(k\). The circuit parity \(P_{k+1}\)'s row \(i\) remains unchanged, while the row \(j\) is updated by performing a bitwise \texttt{XOR} with the row \(i\).

We can use two logical implies (\(\Rightarrow\)) to model the \texttt{XOR} operation. Assuming a CNOT is applied on the directed edge \(\vec{e} = [q_m, q_n]\), 
if \(P_{k, m, i}\) is true, then \(P_{k+1, n, i}\) must differ from \(P_{k, n, i}\); 
otherwise, \(P_{k+1, n, i}\) should remain unchanged. The logical constraints are expressed as follows.

\begin{equation}
\bigvee_{k, \vec{e}, j} \left( \texttt{AND}(cnot_{k, \vec{e}}, P_{k, \vec{e}[0], j}) \Rightarrow P_{k+1, e[1], j} \ne P_{k, \vec{e}[1], j} \right)
\label{eq:cnot_xor_if_true}
\end{equation}

\begin{equation}
\bigvee_{k, \vec{e}, j} \left( \texttt{AND}(cnot_{k, \vec{e}}, \neg P_{k, \vec{e}[0], j}) \Rightarrow P_{k+1, \vec{e}[1], j} = P_{k, \vec{e}[1], j} \right)
\label{eq:cnot_no_change_if_false}
\end{equation}

\begin{equation}
\bigvee_{k, i, j} \left( \texttt{AND} \left(
        \neg cnot_{k, \vec{e}} \mid  q_i 
        \in \vec{e},\; \vec{e} \in \vec{E} \right) \Rightarrow P_{k+1, i, j} = P_{k, i, j} \right)
\label{eq: common_other_rows_unchanged}
\end{equation}

\subsection{Constraints for CNOT Optimal}
For CNOT count optimal, we restrict each \(cnot_{k, e}\) such that at most one CNOT operation can occur across all time steps \(0\leq k<K\). This restriction is enforced by applying \texttt{AtLeast} and \texttt{AtMost} constraints.

\begin{equation}
\bigvee_{k} \left( \texttt{AtLeast}(\{cnot_{k, \vec{e}} \mid \vec{e} \in \vec{E}\},\; 1) \right)
\label{eq:cnot_atleast_once}
\end{equation}

\begin{equation}
\bigvee_{k} \left( \texttt{AtMost}(\{cnot_{k, \vec{e}} \mid \vec{e} \in \vec{E}\}, 1) \right)
\label{eq:cnot_atmost_once}
\end{equation}

\subsection{Constraints for CNOT-Based Doubly Optimal}
 To achieve CNOT-based doubly optimal synthesis, we first minimize the number of CNOT gates to obtain optimal CNOT count. Then, given this optimal CNOT count, we search for the minimal CNOT depth by incrementally adding constraints. To finish the second process, we firstly introduce two auxiliary binary variable matrices:

\begin{itemize}
    \item \textbf{Mapping Gate to Layer:} Define a binary matrix \( D_{k, l} \) where \( 0 \leq k < K \) and \( 0 \leq l < K \). \( D_{k, l} = 1 \) indicates that the CNOT gate \( \textit{cnot}_{k, \vec{e}} \) at time step \( k \) is assigned to layer \( l \).
    
    \item \textbf{Mapping Layer to CNOT Gate:}
    Define another binary matrix \( L_{l, \vec{e}} \), where \( L_{l, \vec{e}} = 1 \) denotes that the CNOT gate acting on edge \( \vec{e} \) appears in layer \( l \).
\end{itemize}

\subsubsection{Valid Layer Assignment} Each CNOT gate must be assigned to exactly one layer, which leads to the following constraints:

\begin{equation}
\bigvee_{k} \texttt{AtLeast}(\{ D_{k,l} \mid 0 \leq l < K \},\; 1)
\label{eq:cnot_depth_atleast_once}
\end{equation}

\begin{equation}
\bigvee_{k} \texttt{AtMost}(\{ D_{k,l} \mid 0 \leq l < K \},\; 1)
\label{eq:cnot_depth_atmost_once}
\end{equation}

\subsubsection{Reduce the Searching Space} To enable efficient search, we enforce temporal continuity within each layer by disallowing alternating or skipping time steps in that layer. That is, if a CNOT gate is assigned to layer \( l \) at time \( k \), then the next gate assigned to layer \( l \) should be at time \( k+1 \) or there is not gates assigned to layer \( l \) after time \(k\). We express this continuity constraint as:

\begin{equation}
\resizebox{\linewidth}{!}{$
\bigvee_{k<K-1, l} \left( D_{k,l} \Rightarrow \texttt{OR}(D_{k+1,l}, \texttt{AND}(\{\neg D_{k+i,l} | 1
\leq i<K-k \} ) \right)
$}
\label{eq: CNOT continuity}
\end{equation}

To avoid permutation search over layer assignments, we constrain the layer indices to increase monotonically with time. That is, if a CNOT gate at time step \(k\) is assigned to layer \(l_k\), and the next gate at \(k+1\) is assigned to \(l_{k+1}\), then \(l_k \leq l_{k+1}\). This constraint is enforced as:

\begin{equation}
\resizebox{\linewidth}{!}{$
\bigvee_{k<K-1, l} \left( D_{k,l} \Rightarrow \texttt{AND}(\{\neg D_{k+i,l-j} | 1
\leq i<K-k;  0
\leq j <l \}  \right)
$}
\label{eq:other_rows_unchanged}
\end{equation}

\subsubsection{Valid Layer Topology}
To ensure that the CNOT gates in each layer conform to the hardware topology, we use \( L_{l,\vec{e}} \) to indicate whether a CNOT gate acting on edge \( \vec{e} \) appears in layer \( l \). We have:

\begin{equation}
\bigvee_{l, \vec{e} \in \vec{E}} \left( L_{l,\vec{e}} = \texttt{OR}(\{ \texttt{AND} (D_{k,l}, \textit{cnot}_{k, \vec{e}}) | 0
\leq k<K \}  \right)
\label{eq:set edges of layer}
\end{equation}

Then, we restrict on each layer there are no two CNOTs interfere on the same
qubit simultaneously:

\begin{equation}
\bigvee_{l,i} \left( 
    \texttt{AtMost}\left(
        \{L_{l, \vec{e}} \mid q_i \in \vec{e} ; \vec{e} \in \vec{E}\},\; 1
    \right)
\right)
\label{eq:not two CNOTs interfere}
\end{equation}

\subsubsection{Control the Depth}

To control the solution circuit adapts depth \(d\), we add the following constraints:

\begin{equation}
\bigvee_{k,\, l \geq d}  \neg D_{k,l}
\label{eq:depth_limit}
\end{equation}

This enforces that no CNOT gate is assigned to a layer index \(l \geq d\). Then the solver is forced to find a solution within the desired depth bound.

\subsection{Constraints for Depth Optimal}

\subsubsection{Valid Layer}

For CNOT depth optimal, we impose two constraints: 1) no two CNOT gates may act on the same physical qubit at the same time step \(k\), and 2) at least one CNOT gate must be scheduled at each time step. These constraints are encoded using \texttt{AtLeast} and \texttt{AtMost} as following:

\begin{equation}
\bigvee_{k} \left( \texttt{AtLeast}(\{cnot_{k, \vec{e}} \mid \vec{e} \in \vec{E}\},\; 1) \right).
\label{eq:cnot limit atleast}
\end{equation}

\begin{equation}
\bigvee_{k,i} \left( 
    \texttt{AtMost}\left(
        \{cnot_{k, \vec{e}} \mid q_i \in \vec{e} ; \vec{e} \in \vec{E}\},\; 1
    \right)
\right)
\label{eq:cnot limit atmost}
\end{equation}

\subsection{Constraints for Depth-Based Doubly Optimal}
For Depth-Based Doubly Optimal, we incrementally add constraints to reduce the total number of CNOTs after obtaining a depth-optimal solution. We begin by restricting the total number of CNOTs to \(n_{\text{cnot}}\) using the following constraint:
\begin{equation}
\texttt{AtMost}\left( \{cnot_{k, \vec{e}} \mid \vec{e} \in \vec{E},\; k \in [0, K] \},\; n_{\text{cnot}} \right).
\label{eq:cnot_atmost_n_{cnot} }
\end{equation}

\subsection{Incremental Solving Algorithm for Doubly Optimal}

To achieve doubly optimal synthesis, we first optimize one metric (either CNOT count or depth), and then introduce additional constraints to optimize the other. The key difference between CNOT-based and depth-based doubly optimal synthesis lies in the specific constraints required. We categorize them into three types of constraints:
\begin{itemize}
  \item \textbf{Optimal Constraints} (\texttt{O\_Constraints}) — used to obtain the primary-optimal solution in Step~1.
  \item \textbf{Doubly-Optimal Constraints} (\texttt{DO\_Constraints}) — used to initialize the secondary optimization in Step~2.
  \item \textbf{Doubly-Control Constraints} (\texttt{DC\_Constraints}) — applied incrementally to optimize the secondary metric while preserving the base-optimal solution.
\end{itemize}

We summarize the types of constraints and the specific constraint used in each synthesis in Table~\ref{tab:doubly_optimal_summary}. Algorithm \ref{alg:incremental_solving_depth} presents our incremental solving procedure for double optimal synthesis.

\begin{algorithm}[H]
\caption{\texttt{HOPPS}(\(I, G, T, \Theta, CP\)): Incremental Solving for Doubly Optimal Circuit}
\begin{algorithmic}[1]
\State \textbf{Input:} \(I\) (initial parity matrix), \(G\) (final parity matrix), \(T\) (parity table), \(\Theta\) (term angles), \(CP\) (coupling map)
\State \textbf{Output:} \(qc\) (optimized quantum circuit)
\For{\(k = n_{\text{term}}\) to \(K_{\max}\)}
    \State \text{Const}\(\gets\)\texttt{O\_Constraints}\((I, G, T, k, CP)\) 
    \State \texttt{model.add(}Const\texttt{)}
    \If{\texttt{model.solve()}}
        \State \(n_{\text{cnot}} \gets \texttt{model.countCNOTs()}\)
        \State \texttt{s\_model} \(\gets\) \texttt{model} 
        \State \text{Const}\(\gets\)\texttt{DO\_Constraints}\((n_\text{cnot})\) 
        \State \texttt{model.add(}\text{Const}\texttt{)} 
        \For{\(n_c = n_{\text{cnot}}\) \textbf{downto} 0}
            \State \text{Const}\(\gets\)\texttt{DC\_Constraints}\((n_{c})\) \State\texttt{model.add(}Const\texttt{)}
            \If{\texttt{model.solve()}}
                \State \texttt{s\_model} \(\gets\) \texttt{model} 
            \Else
                \State \(qc \gets \texttt{s\_model.extract\_circuit}(\Theta)\)
                \State \Return \(qc\)
            \EndIf
        \EndFor
    \EndIf
\EndFor
\end{algorithmic}
\label{alg:incremental_solving_depth}
\end{algorithm}

\begin{table}[htbp]
\caption{Constraints for Doubly Optimal Synthesis}
\begin{center}
\footnotesize
\begin{tabular}{|c|c|c|}
\hline
\textbf{Constraint Type} & \textbf{CNOT-based} & \textbf{Depth-based} \\
\hline
\multirow{2}{*}{\texttt{O\_Constraints}} 
& \multicolumn{2}{c|}{Eq.~\ref{eq:initial_parity}, \ref{eq:final_parity}, \ref{eq:parity_match}, \ref{eq:cnot_xor_if_true}, \ref{eq:cnot_no_change_if_false}, \ref{eq: common_other_rows_unchanged}} \\
\cline{2-3}
& Eq.~\ref{eq:cnot_atleast_once}, \ref{eq:cnot_atmost_once} 
& Eq.~\ref{eq:cnot limit atleast}, \ref{eq:cnot limit atmost} \\
\hline
\texttt{DO\_Constraints} & 
Eq.\ref{eq:cnot_depth_atleast_once}
\ref{eq:cnot_depth_atmost_once}
\ref{eq: CNOT continuity}
\ref{eq:other_rows_unchanged}
\ref{eq:set edges of layer}
\ref{eq:not two CNOTs interfere} & 
None \\
\hline
\texttt{DC\_Constraints} & 
Eq.~\ref{eq:depth_limit} & 
Eq.~ \ref{eq:cnot_atmost_n_{cnot}  } \\
\hline
\multicolumn{3}{l}{\scriptsize Constraints defined in Section~\ref{sec:sat-model}.}
\end{tabular}
\label{tab:doubly_optimal_summary}
\end{center}
\end{table}

\begin{figure*}[ht]
    \centering
    \includegraphics[width=0.85\textwidth]{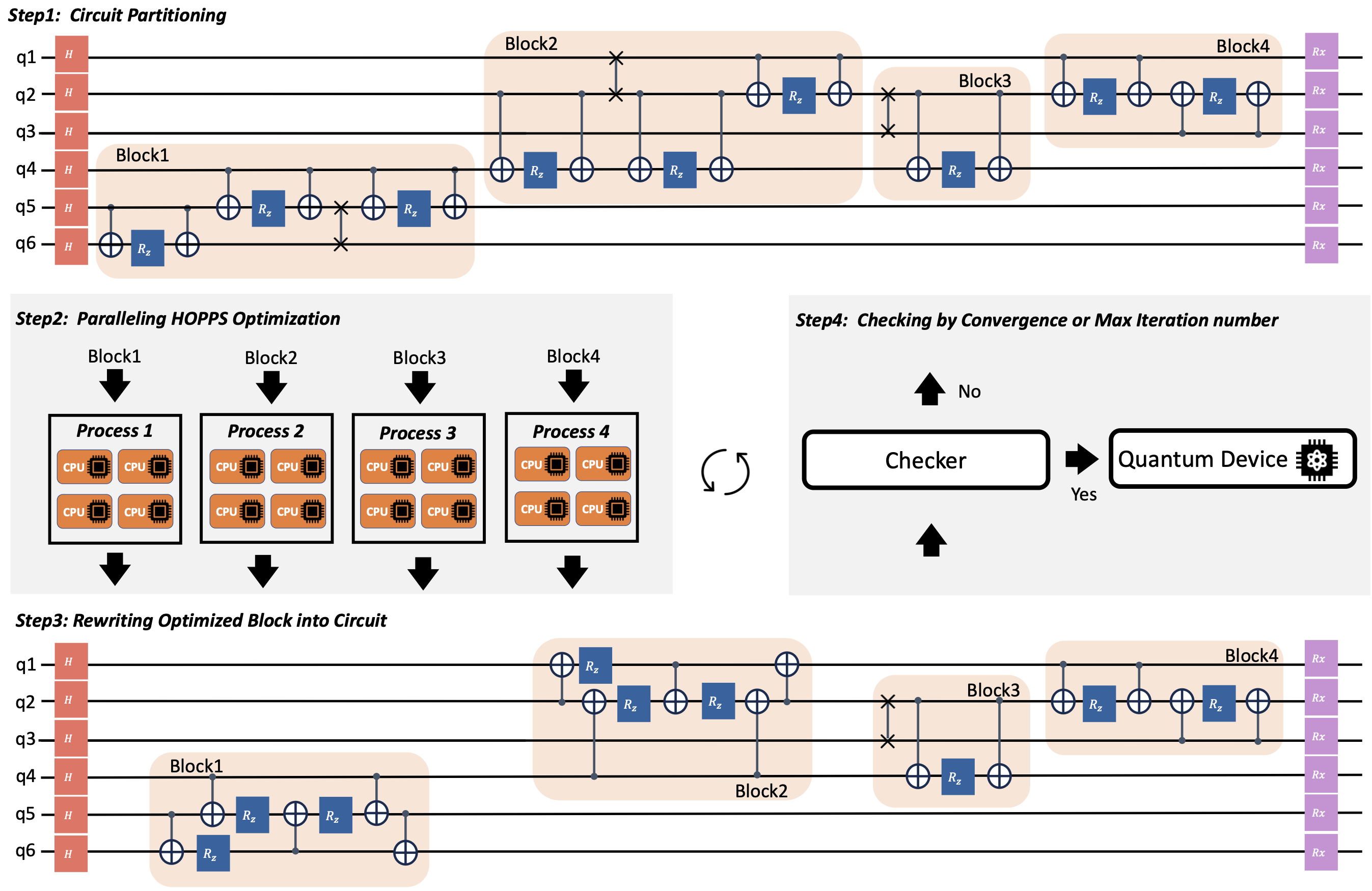}  
    \caption{A pipeline overview of the Blockwise Optimization process: we first partition the circuit into blocks and then apply HOPPS in parallel to optimize each block. Afterward, we track metrics such as CNOT count and CNOT depth to check for convergence. If the circuit has converged, it is compiled for execution on a real device; otherwise, the process is repeated.}
    \label{fig: Blockwise Optimization}
\end{figure*}

\section{HOPPS for Circuit Optimization}
\label{sec: Circuit Optimization}
\subsection{Peephole Optimization}
First, we propose HOPPS can perform peephole optimization for quantum circuit. This idea has been previously explored in works such as~\cite{meuli2018sat, shaik2024optimal2}. Specifically, we identify subcircuits composed solely of \(\{\text{CNOT}, R_z(\theta)\}\) gates as blocks, using HOPPS to synthesize their optimal circuits, and rewrite the original blocks with the optimized versions. HOPPS is flexible in setting different \(T\), \(G\), and \(CP\), making it well-suited for this task.

\subsection{Iterative Blockwise Optimization}
However, some blocks may still face scalability problems while optimizing them, especially in circuits like QAOA, which are almost entirely composed of \(\{\text{CNOT}, R_z(\theta)\}\) gates. To address this situation, we partition large blocks (circuits) into smaller subblocks (subcircuits), each of which is optimized individually and then reassembled into the circuit.

We observe that, sometimes, a single round of optimization does not yield the maximum possible benefit. Therefore, we apply the optimization iteratively until no further improvement can be achieved. We divide this process into two stages. In the first stage, we partition the entire large block into subblocks and optimize each subblock. The process is repeated until convergence, or until the maximum iteration limit is reached. In the second stage, we randomly sample subblocks within larger blocks and apply optimization to them. This stage is repeated for a fixed number of iterations. 
The full procedure is described in the Fig.~\ref{fig: Blockwise Optimization}.

\subsection{Parallelization}

A key advantage of blockwise optimization is that it enables parallel execution of each block independently. Since we are solving a SAT problem, which is inherently suitable for parallelism, we implement a parallel processing strategy where multiple processes are launched, each assigned number of CPUs to enable multithreaded solver execution. Our parallelization approach is illustrated in Step 2 of Fig.~\ref{fig: Blockwise Optimization}.

HOPPS relies on a SAT solver, where computational resources are typically the primary limiting factor rather than memory. For the blockwise optimization, blocks are independent of each other and each block is restricted in size. Thus, the problem requires solving multiple controllable SAT problems in parallel and there are no conflicts between solving different blocks. We therefore consider this a computationally bound problem, and providing more CPUs will help accelerate performance.

\section{Related Work}
\label{sec: Related Work}
We review related work from three perspectives: phase polynomial synthesis; blockwise optimization and synthesis; the use of solvers in quantum compilation.

\textbf{Phase Polynomial Synthesis.}  
Previous studies on phase polynomial synthesis for fault-tolerant circuits focus on minimizing \(T\) gates, as these gates introduce substantial overheads~\cite{amy2019t, ruiz2024quantum, kissinger2020reducing}. 
In the noisy intermediate-scale quantum (NISQ) era, however, CNOT gates become more critical due to their relatively high error rates. Gray-Synth, introduced in~\cite{amy2017cnot}, was originally designed for synthesizing logical circuits. The work by~\cite{shaik2024optimal2} provides optimal solutions for such logical circuit synthesis.  
For physical circuits, heuristic approaches have been proposed for synthesizing dense and large parity tables~\cite{vandaele2022phase, de2020architecture}. Inspired by phase polynomial synthesis,~\cite{dreier2025connectivity} introduced the \emph{twin chain} method to map fully connected circuits under various hardware topologies. Building on this idea,~\cite{montanez2025optimizing} proposed a heuristic extension to handle circuits that are not fully connected. None of the existing research guarantees that phase polynomial synthesis is optimal under arbitrary hardware constraints.

\textbf{Blockwise Optimization and Synthesis.}
Blockwise strategies are widely used for quantum-circuit optimization and synthesis.  
Studies have examined blockwise optimization at the logical-circuit level ~\cite{weiden2022wide,liu2023tackling}.  
Other work focuses on hardware-aware re-optimization that respects device topology~\cite{wu2021reoptimization,wu2020qgo}.  
Nonetheless, runtime and solution quality remain constrained by the efficiency of current resynthesis tools~\cite{smith2023leap,davis2020towards}. Recent work \cite{arora2025local} could optimize both within blocks and across block boundaries.

\textbf{Solver-Based Quantum Compilation.} Quantum compilation encompasses several aspects: circuit synthesis, mapping, and optimization.
For quantum mapping, several studies have focused on using constraint-based methods to achieve optimal mapping. Works such as~\cite{tan2020optimal, lin2023scalable} employ linear integer programming to compute optimal qubit layouts. In contrast,~\cite{shaik2024optimal, jakobsen2025depth, yang2024quantum} leverage SAT solvers to address the same problem. Additionally,~\cite{shaik2025cnot, shaik2024optimal2} presents SAT-based approaches for optimally synthesis Clifford circuits and CNOT circuit respectively.

\begin{table*}[ht]
\scriptsize 
\setlength{\tabcolsep}{3pt}
\renewcommand{\arraystretch}{1.3}
\centering
\caption{Evaluation of HOPPS as Peephole Optimization}
\begin{tabular}{|l|ccc|ccc|ccc|ccc|}
\hline
\textbf{OLSQ \cite{tan2020optimal} Circuit } 
& \multicolumn{3}{c|}{\textbf{CNOTP ~\cite{shaik2024optimal2}}} 
& \multicolumn{3}{c|}{\textbf{Clifford~\cite{shaik2025cnot}}} 
& \multicolumn{3}{c|}{\textbf{HOPPS-CNOT}} 
& \multicolumn{3}{c|}{\textbf{HOPPS-Depth}} \\[-3pt]
\textbf{(Ct$\backslash$Dp)} & \textbf{Ct} & \textbf{Dp} & \textbf{Tm} 
& \textbf{Ct} & \textbf{Dp} & \textbf{Tm} 
& \textbf{Ct} & \textbf{Dp} & \textbf{Tm} 
& \textbf{Ct} & \textbf{Dp} & \textbf{Tm} \\
\hline
MaxCut8 E10 (26$\backslash$16)   & 26 & 16 & 0.2 & 26 & 16 & 2.6 & \multicolumn{3}{c|}{\textit{Out of Limit}} & \textbf{19} & \textbf{9}  & 97.1 \\
MaxCut7 E9  (24$\backslash$18)   & 24 & 18 & 0.2 & 24 & 18 & 3.0 & \multicolumn{3}{c|}{\textit{Out of Limit}} & \textbf{18} & \textbf{8}  & 21.3 \\
MaxCut6 E7  (14$\backslash$14)   & 14 & 14 & 0.1 & 14 & 14 & 0.2 & 14 & \textbf{6}  & 36560.6 & 14 & \textbf{6}  & 3.0 \\
MaxCut5 E7  (20$\backslash$13)   & 20 & 13 & 0.5 & 20 & 13 & 0.5 & \textbf{14} & 11 & 33.0    & 15 & \textbf{8}  & 3.5 \\
LABS4        (8$\backslash$7)    &  8 &  7 & 0.2 &  8 &  7 & 0.4 & \textbf{6}  & \textbf{5}  & 0.2     &  \textbf{6} & \textbf{5}  & 0.1 \\
LABS5       (44$\backslash$35)   & 35 & 28 & 2.2 & 35 & 28 & 2.4 & \textbf{22} & 21 & 1977.7   & 23 & \textbf{15} & 47.7 \\
barencotof3 (44$\backslash$41)   & 38 & 36 & 1.5 & 37 & 32 & 8.2 & \textbf{28} & 28 & 2.5      & 29 & \textbf{26} & 1.9 \\
barencotof4 (78$\backslash$70)   & 74 & 65 & 3.4 & 69 & 62 & 14.0 & \textbf{59} & 49 & 11.7     & 60 & \textbf{45} & 5.4 \\
barencotof5(110$\backslash$95)   &108 & 94 & 5.1 & 97 & 79 & 24.0 & \textbf{88} & 75 & 19.8     & 89 & \textbf{68} & 8.6 \\
mod54       (56$\backslash$49)   & 48 & 40 & 2.3 & 39 & 33 & 25.6 & \textbf{30} & 28 & 102.0    & 32 & \textbf{24} & 15.5 \\
modmult55  (131$\backslash$79)   &115 & 69 &13.0 &125 & 75 &603.7 & \textbf{106}& 63 & 458.9    &110 & \textbf{59} & 16.8 \\
qft4       (105$\backslash$101)  & 87 & 82 & 5.1 & 80 & 75 & 16.2 & \textbf{72} & 69 & 9.9      & 73 & \textbf{60} & 10.9 \\
rcadder6   (145$\backslash$102)  &139 &100 &12.1 &128 & 96 &445.9 & \textbf{132}& 90 & 15.8     &\textbf{132} & \textbf{89} & 8.9 \\
tof3        (34$\backslash$33)   & 34 & 31 & 1.4 & 34 & 33 & 3.9 & \textbf{23} & 19 & 2.2      & 24 & \textbf{17} & 1.5 \\
tof4        (61$\backslash$55)   & 61 & 53 & 3.6 & 55 & 52 & 13.6 & \textbf{53} & 41 & 27.0     & 55 & \textbf{37} & 6.0 \\
tof5        (80$\backslash$71)   & 76 & 68 & 3.3 & 69 & 58 & 18.5 & \textbf{68} & 61 & 10.6     & 72 & \textbf{53} & 5.0 \\
vbeadder3   (86$\backslash$75)   & 79 & 67 & 4.3 & 72 & 55 & 100.8 & \textbf{62} & 43 & 25.3     & 66 & \textbf{39} & 21.3 \\
\cline{1-13}
\textbf{Max (\%)}  & 20.4 & 20.0 &      &  30.3 & 32.6 &  & \textbf{50.0} & \textbf{57.1} &      & 47.4 & \textbf{57.1} &      \\
\textbf{Mean (\%)} & 5.9 & 6.8 &      & 10.0  & 11.1 &  &  \textbf{23.3} & 30.3 &      & 22.3 & \textbf{39.9} &      \\
\hline
\end{tabular}
\label{tab:peephole_comparing}
\end{table*}

\section{Experiment}
\label{sec:experiments}
\subsection{Experiment Setup}
\subsubsection*{\textbf{Backend}}

In this paper we focus on widely used and publicly accessible quantum hardware provided by IBM. We target the large device \texttt{ibm\_kyiv}, as well as the small device \texttt{ibmq\_melbourne}. The \texttt{ibm\_kyiv} device can be accessed freely through IBM.

\subsubsection*{\textbf{Benchmark}}

For general optimal hardware-mapped circuits, we use benchmark circuits from~\cite{shaik2024optimal}, which were optimally permutation-aware mapped onto ibmq\_melbourne.

For QAOA-based benchmarks, we use circuits for two types of combinatorial problems: MaxCut and LABS (Low Autocorrelation Binary Sequences), 2-local and 4-local QAOA ansätze. For MaxCut we generate random graphs and regular graphs using the \texttt{networkx} library. The label ``MaxCut n R m'' refers to an \(n\)-qubit QAOA circuit for a regular graph of degree \(m\), while ``MaxCut n E m'' refers to an \(n\)-qubit QAOA circuit for a random graph with \(m\) edges. For the LABS problem we use instances from~\cite{shaydulin2024evidence-LABS}.

\subsubsection*{\textbf{Implementation}}
All experiments were conducted by using Python, with the Z3 solver used for SAT. For iterative blockwise optimization, we use the \texttt{QuickPartitioner} provided in BQSKit~\cite{osti_1785933}.

The experiments were conducted on an HPC system equipped with Intel(R) Xeon(R) Silver 4216 CPUs @ 2.10\,GHz. For the HOPPS versus QSearch comparison, each job was executed on a single compute node with 8 CPUs and 128\,GB of RAM. The optimization experiments were run on a single compute node with 32 CPUs and 128\,GB of RAM. Parallel tasks were executed using up to 8 logical threads via Python’s \texttt{multiprocessing} module.

\subsubsection*{\textbf{Comparison}}
First, HOPPS is a circuit synthesis tool; therefore, we compare it with QSearch\cite{davis2020towards} from BQSKit, another hardware-aware circuit synthesis tool that provides near-optimal solutions. Due to their result is close to optimal, we maintain the same performance in terms of CNOT count and depth, and focus primarily on runtime comparison.

Second, HOPPS supports peephole optimization. We use OLSQ~\cite{lin2023scalable}, an optimal layout synthesis method, as the baseline. For comparison, we apply two optimal-synthesis–based peephole methods: CNOTP~\cite{shaik2024optimal2} and Clifford~\cite{shaik2025cnot}.

Third, for large-scale \( \{\text{CNOT}, R_z(\theta)\} \) circuits, we compare our approach with BQSKit’s partition-and-resynthesis strategy, specifically using the \texttt{QuickPartitioner} combined with the \texttt{QSearchSynthesisPass}. The baseline compilers include Qiskit, 2QAN~\cite{lao20222qan}, and ArPhase~\cite{de2020architecture}. Among these methods, 2QAN is limited to two-local QAOA circuits only, whereas ArPhase excels on multi-local circuits but is less effective for two-local instances. Therefore, we would not have 2QAN results for the LABS category, and ArPhase is not reported for the larger MaxCut benchmarks.

\subsubsection*{\textbf{Metrics}}

To compare circuit sizes, we use commonly adopted metrics: \textbf{CNOT count} (Ct), which refers to the number of CNOT gates in the circuit, and \textbf{CNOT depth} (Dp), defined as the length of the longest path consisting only of CNOT gates. We focus on CNOT depth because, in current quantum devices, two-qubit gates typically have longer durations than do single-qubit gates. A circuit with many single-qubit gates along the critical path should not be considered equivalent in duration to one with many two-qubit gates. Therefore, we argue that CNOT depth is a more appropriate metric for evaluating circuit depth. We compute the improvement ratio as \[\frac{\text{baseline} - \text{our method}}{ \text{baseline}}\] which reflects the percentage of improvement.
To evaluate algorithm efficiency, we also report the \textbf{solving time} (Tm) (in seconds).

\subsection{HOPPS as a \(\{\mathrm{CNOT}, R_z(\theta)\}\) Peephole Optimization Pass}

HOPPS can be applied as a peephole optimization pass on mapped circuits. We use various types of circuits mapped by OLSQ on \texttt{ibmq\_melbourne} as the baseline. We extract all \( \{\text{CNOT}, R_z(\theta)\} \) blocks and perform resynthesis on each block. For QAOA circuits, which are almost entirely composed of \( \{\text{CNOT}, R_z(\theta)\} \) operations, HOPPS synthesizes the entire circuit directly. 
For comparisons, we compare against CNOTP~\cite{shaik2024optimal2}---another peephole optimization method focused on resynthesizing CNOT blocks---and Clifford~\cite{shaik2025cnot}, a peephole optimization method focused on resynthesizing Clifford blocks. The result is presented in the Table~\ref{tab:peephole_comparing}.

Compared with , HOPPS-CNOT reduces the CNOT count up to \(\textbf{50.0\%}\) and the CNOT depth by \(\textbf{57.1\%}\). 
HOPPS-Depth achieves average reductions of \(\textbf{23.3\%}\) in CNOT count and \(\textbf{57.1}\%\) in CNOT depth. 
Across all benchmark circuits, both HOPPS variants outperform CNOTP and Clifford.

\begin{table*}[ht]
\scriptsize
\setlength{\tabcolsep}{3pt}
\renewcommand{\arraystretch}{1.3}
\centering
\caption{Evaluation of Iterative Blockwise Optimization Strategy}
\begin{tabular}{|l|l|ccc|ccc|ccc|ccc|ccc|}
\hline
\textbf{Compiler} & \textbf{Circuit} 
& \multicolumn{3}{c|}{\textbf{BQSKit (3)}} 
& \multicolumn{3}{c|}{\textbf{HOPPS-CNOT (3)}} 
& \multicolumn{3}{c|}{\textbf{HOPPS-Depth (3)}} 
& \multicolumn{3}{c|}{\textbf{HOPPS-CNOT (5)}} 
& \multicolumn{3}{c|}{\textbf{HOPPS-Depth (5)}} \\
& (Ct$\backslash$Dp)
& \textbf{Ct} & \textbf{Dp} & \textbf{Tm} 
& \textbf{Ct} & \textbf{Dp} & \textbf{Tm} 
& \textbf{Ct} & \textbf{Dp} & \textbf{Tm}
& \textbf{Ct} & \textbf{Dp} & \textbf{Tm}
& \textbf{Ct} & \textbf{Dp} & \textbf{Tm} \\
\hline
\multirow{9}{*}{Qiskit} 
& MaxCut16 R4 (137$\backslash$79)   & 127  & 74  & 173.8 & 119 & 74 & 21.8 & 121 & 73 & 20.5 & \textbf{111}  & 60  & 329.4 & \textbf{117} & 62 & 35.8 \\
& MaxCut32 R4  (392$\backslash$193) &   389   & 183    &   176.3     & 375&185 & 41.3&372 &177 &34.9 & \textbf{354}  & 159 & 687.3 & 354 & \textbf{146} & 65.4 \\
& MaxCut64 R4 (1358$\backslash$561)  &  1285    &   549  & 374.7       &1286 &528 &79.3 &1267 &537 &73.1 & \textbf{1185} & \textbf{465} & 13460.6 & 1214 & 467 & 200.4 \\
& MaxCut96 R4 (2649$\backslash$1086)  &   2386   &  891   &  7086.1  & 2378    & 902 & 421.0  & 2385 & 878 & 206.1 & 2316 &  800   & 78542.4 & \textbf{2288}    & \textbf{745}   & 753.4 \\
& LABS20 (12585$\backslash$7459)  &  10183   & 5826    &  1224.0      & 10079 & 5683 & 768.9 &10041 &5629 &529.0 & \textbf{8522} & 4245 & 12849.0 & 8323 & \textbf{4057} & 695.5 \\
& LABS15  (4128$\backslash$2808)   & 3234    &  2176   &  496.7      &3171 &2085 &248.9 & 3196 & 2112 & 187.5& \textbf{2594} & 1490  & 18018.5 & 2608 & \textbf{1469} & 610.5 \\
& LABS10   (875$\backslash$672)     &  663   &  513   &   169.7     &637 & 470 & 67.8 & 641 & 473 & 55.2 & \textbf{486}  & 311  & 5476.1  & 506 & \textbf{318} & 46.04 \\
\cline{2-17}
& \textbf{Max (\%)}    & 24.2    &   23.6  &        & 27.2 &30.0 & & 26.7 & 29.6 & & \textbf{44.4} & 42.1 &        & 37.5 & \textbf{42.4} &        \\
& \textbf{Mean (\%)}   &  12.6   &   14.2  &        & 14.7 & 16.0 & & 14.7 & 16.8 & & \textbf{23.9} & \textbf{23.0} &        & 19.2 & 22.1 &        \\
\hline

\multirow{6}{*}{2QAN \cite{lao20222qan}} 
& MaxCut16 R4 (168$\backslash$68)  & 159    &  64   &   83.6     &158 &63 &19.1 &157 &65 & 16.4 & \textbf{148}  & \textbf{60}  & 2036.4 & 152 & 71 & 65.8 \\
& MaxCut32 R4  (437$\backslash$97)  &  411   &  97   & 148.9   & 403 & 93 & 39.6 &403 & 96& 21.4 & \textbf{385}  & 94  & 2433.1 & 387 & \textbf{91} & 71.5 \\
& MaxCut64 R4  (1169$\backslash$209)  &  1103   &  208   &    191.4    & 1097 & 205 & 78.7 & 1108 &201 & 54.6 & \textbf{1031} & \textbf{171} & 25417.8 & 1041 & 184 & 232.91 \\
& MaxCut96 R4  (2205$\backslash$262)   &  2083   &  262   &   5928.2    & 2078 & 258 & 166.4 & 2063 & 254 & 102.1 & 1962   & 274  & 62838.1        & \textbf{1952}    & \textbf{239}   & 444.1 \\
\cline{2-17}
& \textbf{Max (\%)}   &  5.9   & 5.8    &        & 7.7 & 7.3 & & 7.7 &4.4 & & \textbf{11.9} & \textbf{18.1} &        & 11.4 & 11.9 &        \\
& \textbf{Mean (\%)}  &   5.6  &   1.5  &        & 6.4 & 3.7 & & 6.4 &3.0 & & \textbf{11.6} & \textbf{7.1} &        & 10.8 & 5.6 &        \\
\hline

\multirow{7}{*}{ArPhase \cite{de2020architecture}} 
& MaxCut16 R4 (463$\backslash$274)   & 461    &  273   &     73.2   & 461 & 272 &  17.8 & 461 & 272 & 18.0 & \textbf{448}  & 266 & 4166.8 & 450 & \textbf{264} & 58.0 \\
& MaxCut32 R4 (1949$\backslash$679)  & 1942    & 685    &     41.6   & 1938 & 678 & 154.5 &1943 & 681 & 17.8 & \textbf{1919} & \textbf{680} & 155.8  & 1922 & 683 & 98.9 \\
& LABS20  (7218$\backslash$2380)   & 7210    &  2384   &  290.1      & 7211 & 2384 &236.9 & 7211 & 2383 & 135.9 & \textbf{7107} & 2375 & 4749.6 & 7130 & \textbf{2359} & 303.1 \\
& LABS15   (2229$\backslash$982)   &  2207   & 983    & 167.7  & 2213 & 988 & 107.2 & 2214 & 983 & 74.1 & \textbf{2146} & 977  & 2611.7 & 2151 & \textbf{971} & 122.3 \\
& LABS10    (363$\backslash$220)   &  360   &  216   &  111.5 & 360 & 216 & 24.3 & 360 & 217 & 27.5 & \textbf{339}  & 204  & 461.1  & 342 & \textbf{200} & 105.8 \\
\cline{2-17}
& \textbf{Max (\%)}     &  0.9   & 1.8    &        & 0.8 & 1.8 & &0.8 & 1.3 & & \textbf{6.6} & 7.2 &        & 5.7 & \textbf{9.0} &        \\
& \textbf{Mean (\%)}     &  0.5   & 0.2    &        & 0.5 & 0.3 & & 0.4 & 0.3 & & \textbf{3.3} & \textbf{2.1} &        & 2.9 & 2.0 &        \\
\hline
\multicolumn{17}{l}{\scriptsize “BQSKit (3)” indicates that the circuit is partitioned into 3 qubit blocks using the BQSKit strategy, the same notation is used for the other methods.}
\end{tabular}
\label{tab:optimal_comparing}
\end{table*}

\subsection{HOPPS as an Iterative Block Optimization Strategy}

For large \( \{ \text{CNOT}, R_z(\theta) \} \) blocks, we propose an iterative block optimization strategy. We use various types of circuits compiled by Qiskit, 2QAN~\cite{lao20222qan} and ArPhase~\cite{de2020architecture} on \texttt{ibmq\_kyiv} as the baseline. For comparison, we use an iterative version of BQSKit’s partition-and-resynthesis approach, specifically combining the \texttt{QuickPartitioner} with the \texttt{QSearchSynthesisPass} in multiple iterations. Since QSearch does not scale beyond 3 qubits, the BQSKit strategy is applied using 3-qubit block partitions. To ensure a fair comparison, we evaluate HOPPS with 3-qubit block partitions and additionally with 5-qubit block partitions to demonstrate its scalability. In some cases, a block becomes too deep; therefore, we cap the maximum block depth at 20. For each circuit, we perform 5 iterations of full-circuit partitioning, followed by 5 iterations sampling blocks. Iterations can be stopped early if no explicit improvement is achieved or out of 24 hours. The result is shown in the Table~\ref{tab:optimal_comparing}.

Using the iterative optimization strategy with HOPPS, we achieving reductions in CNOT count and depth of up to \(\textbf{44.4\%}\) and \(\textbf{42.4\%}\) relative to Qiskit, \(\textbf{11.9\%}\) and \(\textbf{18.1\%}\) relative to 2QAN, and \(\textbf{6.6\%}\) and \(\textbf{9.0\%}\) relative to ArPhase, respectively. Moreover, the iterative optimization strategy that integrates HOPPS surpasses the one built with BQSKit.

\begin{figure}[htbp]
\centering
\begin{subfigure}[b]{\linewidth}
\centering
\includegraphics[width=\linewidth]{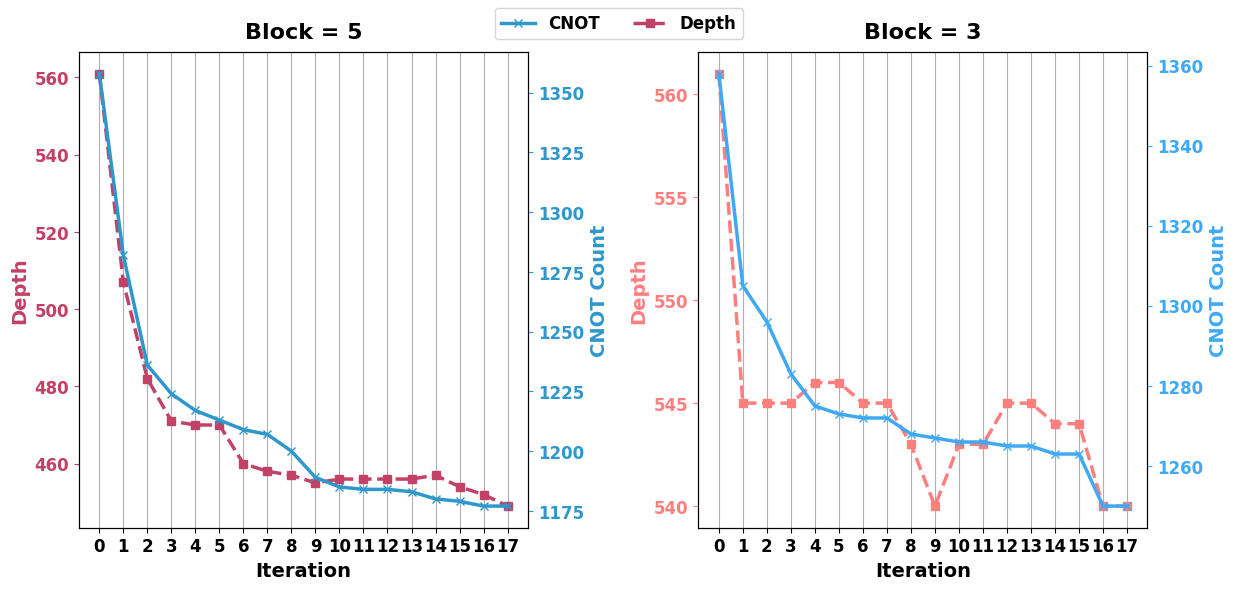}
\caption{Iteration of CNOT count and CNOT depth for the iterative blockwise optimization on the Qiskit MaxCut 64R4 circuit, optimized by HOPPS-CNOT with different block sizes.  We perform 10 iterations of full-circuit partitioning,
and 10 iterations sampling blocks.}
\label{subfig:iteration3}
\end{subfigure}

\vspace{1em} 

\begin{subfigure}[b]{\linewidth}
    \centering
    \includegraphics[width=\linewidth]{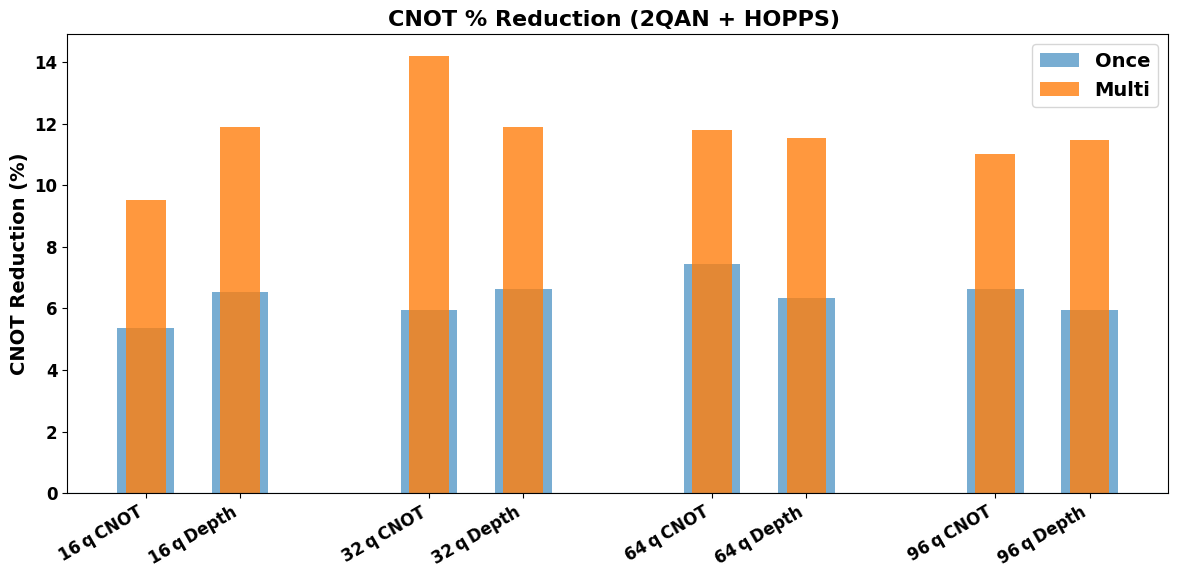}
    \caption{CNOT reduction rate for multi-pass versus single-pass iterations across 2QAN mapped MaxCut R4-QAOA circuits of varying qubit counts.}
    \label{subfig:once_multi}
\end{subfigure}

\caption{Behavior of Iterative Blockwise Optimization Strategy.}
\label{fig:iteration_comparison}

\end{figure}

\begin{figure}[htbp]
    \centering
    \includegraphics[width=\linewidth]{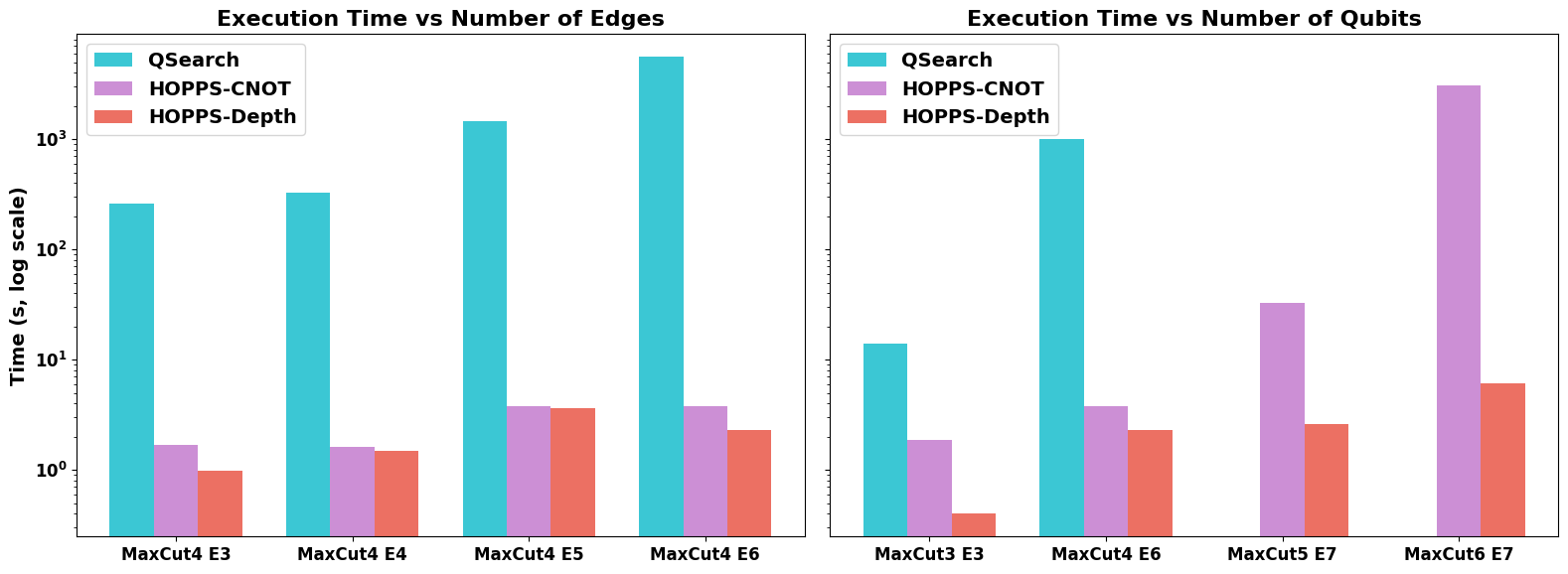}
   \caption{HOPPS vs. QSearch \cite{davis2020towards} Synthesis Time.
    Under equivalent synthesis quality—defined as achieving the same CNOT count and depth—HOPPS requires less execution time compared to QSearch. In the left figure, we fix the number of qubits and vary the number of edges to examine how synthesis time scales with circuit depth. In the right figure, we maintain a similar number of edges while increasing the number of qubits to evaluate scalability with circuit width. The results show that HOPPS is more scalable, particularly depth HOPPS. This demonstrates the advantage of HOPPS for synthesizing \(\{ \text{CNOT}, R_z(\theta) \}\) circuits.
    }

    \label{fig:HOPPS vs QSearch}
\end{figure}

We also provide examples that illustrate the benefit of iterative optimization. In Fig.~\ref{subfig:iteration3}, increasing the number of iterations steadily reduces both the CNOT count and the CNOT depth until they converge. In Fig.~\ref{subfig:once_multi}, circuits of various sizes achieve larger reductions when optimized multiple times rather than just once.

In conclusion, the iterative blockwise optimization strategy is better than perform single-pass, achieving substantial improvements over Qiskit and 2QAN and a slight improvement over ArPhase. This is because ArPhase targets dense parity tables, while the others are SWAP-based compilers. SWAP-based compilers get more chance to optimized by HOPPS.

\subsection{Runtime Comparison: HOPPS vs. QSearch}
To evaluate HOPPS’s efficiency, we compare its runtime against QSearch in BQSKit. We use small circuits that allow the synthesizer to synthesize the entire circuit directly. These circuits are chosen so that both methods yield identical CNOT counts and depths, ensuring optimality. If the runtime exceeds 2 hours, the result is omitted from the figure. We explore both the runtime scale with the circuit depth and width, and the results are presented in Fig.~\ref{fig:HOPPS vs QSearch}. HOPPS demonstrates better scalability than QSearch with respect to both depth and width.

\section{Discussion}
\label{sec: Discussion}

In this section, we highlight several advantages of HOPPS.

\textbf{Efficiency for QAOA:} In QAOA algorithms, parameters are iteratively updated across optimization steps. With methods such as~\cite{weiden2022wide, arora2025local}, each parameter update requires recompilation and resynthesis of the circuit, which incurs significant overhead. In contrast, HOPPS allows the QAOA circuit to be compiled only once. Subsequent parameter updates can be applied directly to the circuit without recompilation, significantly reducing overall compilation time.

\textbf{Optimality Checking:} After mapping or applying other optimizations to a circuit, it is often beneficial to check whether the \( \{ \text{CNOT}, R_z(\theta) \} \) subcircuit is optimal. This type of optimality check is not addressed by prior work such as~\cite{de2020architecture, vandaele2022phase}. HOPPS provides a way to verify and improve such blocks, potentially enhancing circuit quality even after traditional compilation flows.

\textbf{Interoperability.} While this paper adopts BQSKit for circuit partitioning, we claim HOPPS can be suitably integrated with other frameworks, such as those proposed in \cite{weiden2022wide, ping2025high}.

\section{Conclusion and Future Work}
\label{sec: Conclusion}

In this paper, we proposed \textbf{HOPPS}: a \emph{Hardware-Aware Optimal Phase Polynomial Synthesis} tool. HOPPS can synthesize \(\{ \text{CNOT}, R_z \}\) circuits under arbitrary device topologies and provides both CNOT-based and depth-based doubly optimal results (as discussed in Section~\ref{subsec: Optimality}). Compared to QSearch, HOPPS achieves better runtime performance and scales more effectively with circuit width and depth.

We further demonstrated that HOPPS can serve as a quantum circuit optimizer. For small blocks, it performs direct synthesis; for large blocks—where optimal synthesis becomes intractable—we introduced an iterative blockwise optimization strategy. We show that modern compilers such as ~\cite{tan2020optimal}, Qiskit, 2QAN~\cite{lao20222qan}, and ArPhase~\cite{de2020architecture} can all benefit from HOPPS optimization, outperforming existing optimization strategies. Beyond circuit optimization, HOPPS offers several broader applications:

\textbf{Template discovery}—HOPPS can uncover optimization templates that are difficult for humans to identify manually.

\textbf{Hardware-aware template design}—It can guide the creation of specialized templates for particular topologies, such as QAOA circuits mapped to the \emph{twin chain} architecture~\cite{dreier2025connectivity}.

\textbf{Dataset generation for AI models}—HOPPS produces high-quality synthesized circuits that can serve as training data for AI-driven synthesis and optimization tools.

In future work, we plan to integrate HOPPS into end-to-end quantum-compilation pipelines and explore its synergy with learning-based methods.

\section*{Acknowledgment}
This work was supported in part by the NSF research grant 2216923.
This material is based upon work supported by the DOE-SC Office of Advanced Scientific Computing Research MACH-Q project under contract number DE-AC02-06CH11357.

\bibliographystyle{IEEEtran}
\bibliography{sample_base}

\end{document}